**An Analysis of Apparent r-Mode Oscillations in Solar Activity, the Solar Diameter, the Solar Neutrino Flux, and Nuclear Decay Rates, with Implications Concerning the Sun's Internal Structure and Rotation, and Neutrino Processes.**

P.A. Sturrock[a,*], L. Bertello[b], E. Fischbach[c], D. Javorsek II[d], J.H. Jenkins[ce], A. Kosovichev[f], A.G. Parkhomov[g]

[a] Center for Space Science and Astrophysics, Stanford University, Stanford, CA 94305-4060, USA
[b] National Solar Observatory, 950 North Cherry Avenue, Tucson, AZ 85719
[c] Department of Physics, Purdue University, West Lafayette, IN 47907, USA
[d] Air Force Flight Test Center, Edwards AFB, CA 93524, USA
[e] School of Nuclear Engineering, Purdue University, West Lafayette, IN 47907
[f] W. W. Hansen Experimental Physics Laboratory, Stanford University, Stanford, CA 94305, USA
[g] Institute for Time Nature Explorations, Lomonosov Moscow State University, Moscow, Russia

* Corresponding author. Tel +1 6507231438; fax +1 6507234840.
Email address: sturrock@stanford.edu

Luca Bertello, bertello@noao.edu
Ephraim Fischbach, Ephraim@physics.purdue.edu
Daniel Javorsek II, javorsek@hotmail.com
Jere Jenkins, jere@purdue.edu
Alexander Kosovichev, sasha@quake.stanford.edu
Alex Parkhomov, alexparh@mail.ru

This article presents a comparative analysis of solar activity data, Mt Wilson diameter data, Super-Kamiokande solar neutrino data, and nuclear decay data acquired at the Lomonosov Moscow State University (LMSU). We propose that salient periodicities in all of these datasets may be attributed to r-mode oscillations. Periodicities in the solar activity data and in Super-Kamiokande solar neutrino data may be attributed to r-mode oscillations in the known tachocline, with normalized radius in the range 0.66 to 0.74, where the sidereal rotation rate is in the range 13.7 to 14.6 year$^{-1}$. We propose that periodicities in the Mt Wilson and LMSU data may be attributed to similar r-mode oscillations where the sidereal rotation rate is approximately 12.0 year$^{-1}$, which we attribute to a hypothetical "inner" tachocline separating a slowly rotating core from the radiative zone. We also discuss the possible role of the RSFP (Resonant Spin Flavor Precession) process, which leads to estimates of the neutrino magnetic moment and of the magnetic field strength in or near the solar core.

## 1. Introduction

One of the many puzzles of solar physics is the process responsible for the Rieger oscillation [1] and related oscillations. This oscillation, with a period of about 153 days, was first discovered in hard X-ray data acquired by the Gamma Ray Spectrometer on the *Solar Maximum Mission*. Several similar (typically intermittent) oscillations, with periods of months or years, have subsequently been discovered in a variety of solar data. Bai has identified eight oscillations with periods of 33.5, 51, 63, 76, 84, 128, 153, and 257 days, noting that several (but not all) of these periods are multiples of 25.5 days, leading him to suggest that the Sun has an internal clock with this period [2]. However, the oscillations with periods of 33.5, 63, and 84 days do not conform to that pattern.

Lou has shown that oscillations with periods 51, 76, 128, and 153 days may have their origin in equatorially trapped Rossby waves and mixed Rossby-Poincare waves in a magnetized surface layer [3]. However, that model does not seem to account for the oscillations with periods of 33.5, 63, 84, or 257 days.

Zaqarashvili et al. have shown that the Rieger oscillation may be attributed to a magnetic Rossby wave that has its origin in the combined effect of latitudinal differential rotation and magnetic field [4]. However, their model does not address the other seven oscillations identified by Bai.

In this article, we point out that all eight of the periodicities identified by Bai may be attributed to r-mode oscillations [5] in the known solar tachocline with a normalized radius of approximately 0.7 [6]. We also point out that we have found a closely similar set of oscillations (but with systematically longer periods) in Mt Wilson diameter measurements [7] and in nuclear decay data acquired at the Lomonosov Moscow State University (LMSU) [8]. We suggest that these oscillations (in diameter and decay data) also are r-mode oscillations, originating not in the known tachocline, but in a hypothetical second "inner" tachocline that separates a slowly rotating core from the radiative zone at a normalized radius probably in the range 0.2 to 0.3.

We present the basic equations for r-mode oscillations in Section 2, and we discuss the attribution of the Rieger and related periodicities in solar activity data to r-mode oscillations in the known (outer) tachocline in Section 3. In Section 4, we show that certain periodicities in the Super-Kamiokande solar neutrino data may be attributed to r-mode oscillations in the same region. In Section 5, we discuss periodicities in the Mt Wilson diameter measurement [7], and in Section 6 we discuss similar periodicities in nuclear decay data acquired at the Lomonosov Moscow State University (LMSU) [8].

We discuss the possible role of neutrinos and the Resonant Spin Flavor Precession (RSFP) process in Section 7, and we present further discussion in Section 8. In Appendix A, we discuss the possibility that r-mode oscillations may be excited by a Kelvin-Helmholtz instability due to a radial gradient in angular velocity. In Appendix B, we review comments that have been made concerning evidence for the variability of nuclear decay rates.

## 2. r-Mode Oscillations

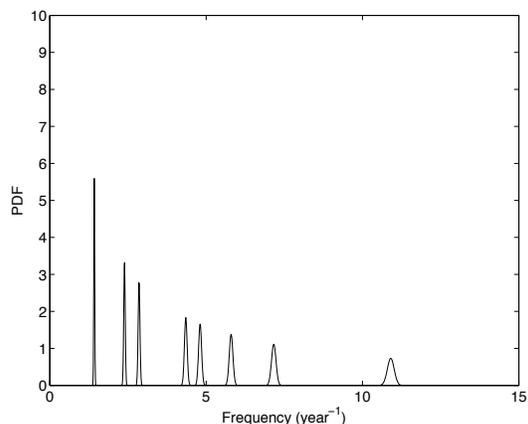

Figure 1. Representation of the oscillation frequencies identified by Bai [2]

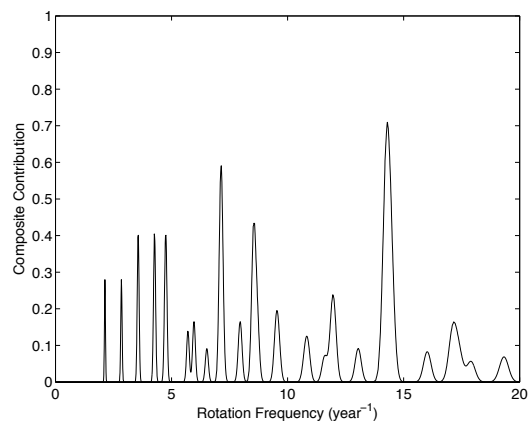

Figure 2. Composite contribution to the r-mode oscillation indicator. The principal peak is at 14.30 year$^{-1}$.

We first consider r-mode oscillations as they occur in a uniform and uniformly rotating fluid sphere with sidereal rotation frequency $\nu_R$. All frequencies will be measured in cycles per year. Since r-mode oscillations are retrograde, the "absolute" frequency of an r-mode oscillation as measured in an inertial reference frame is given, to good approximation, by

$$\nu_A(l,m) = m\nu_R - \frac{2m\nu_R}{l(l+1)}, \qquad (1)$$

where $l$ and $m$ are two of the three familiar spherical harmonic indices [5]. The allowed values of $l$ and $m$ are $l = 2, 3, ..., m = 1, ..., l$. Since this frequency does not depend on the radial index $n$, and since a thin spherical-shell wave function may be decomposed into a set of spherical harmonics with different *n* values, we may infer that Equation (1) also gives the frequency of r-mode oscillations confined to a thin spherical shell. As measured by an observer on Earth, the oscillation frequency will be given by

$$\nu_E(l,m) = m(\nu_R - 1) - \frac{2m\nu_R}{l(l+1)}. \tag{2}$$

We now consider the possibility that the r-mode oscillation interacts with some structure (such as a magnetic flux tube) that rotates with the Sun. An arbitrary structure may be regarded as a superposition of magnetic-field configurations with various values of the longitudinal index $m_S$. The interplay of an r-mode with the cylindrically symmetrical component (with $m_S = 0$) will present the same time-dependence as the r-mode itself, as given by Equation (2).

More generally, the interplay of an r-mode oscillation with a magnetic-field component that has a periodicity index $m_S$ (which can have either sign) will lead to oscillations that, as seen from Earth, would have the frequency

$$\nu_S(l,m) = m_S(\nu_R - 1) - \nu_E(l,m). \tag{3}$$

The case $m_S = m$ is particularly interesting, since it would lead to low-frequency oscillations with frequencies given by

$$\nu_S(l,m) = \frac{2m\nu_R}{l(l+1)}. \tag{4}$$

his is, of course, the r-mode oscillation frequency that would be measured by an observer co-moving with that region of the solar interior. It appears that, for reasons yet to be explored, in general these oscillations seem to have a more pronounced influence on observational quantities than those corresponding to other values of $m_S$, perhaps simply because they have lower frequencies. However, we shall see in Section 4 that, in Super-Kamiokande solar neutrino data, E-type oscillations are more significant than S-type oscillations.

### 3. Solar Activity Oscillations

We now examine the possibility that the Rieger-type oscillations, such as those identified by Bai [2], may be attributed to oscillations with frequencies given by Equations (2) and (4). If we assume that each of the detected periodicities may be characterized by a central frequency $\nu_k$ and an uncertainty $\Delta\nu_k$, then the probability distribution function for the *k*'th periodicity is given by

$$P_k(\nu) = \frac{1}{(2\pi)^{1/2} \Delta\nu_k} \exp\left(-\frac{1}{2}\left(\frac{\nu - \nu_k}{\Delta\nu_k}\right)^2\right). \tag{5}$$

On the assumption that the uncertainty in the frequency assigned to an oscillation, on the basis of observational data, is likely to be proportional to the frequency itself, we adopt the approximation that the values of $\Delta\nu_k$ are proportional to $\nu_k$, and adopt $\Delta\nu_k = \nu_k/Q$. Figure 1 shows the sum of these curves,

$$F(\nu) = \sum_k P_k(\nu), \tag{6}$$

for Q = 100. (The results of our calculations prove not to be sensitive to the assumed value of Q.)

We now wish to find the sidereal rotation frequency that gives the best fit between the data summarized in Equation (6) and the r-mode frequencies given by Equation (4). Focusing on oscillations corresponding to $l = 2, m = 1$, $l = 2, m = 2$, $l = 3, m = 1$, $l = 3, m = 2$, $l = 3, m = 3$, $l = 4, m = 1$, $l = 4, m = 2$, $l = 4, m = 3$, and $l = 4, m = 4$, we carry out the comparison by forming the sum

$$H(v_R) = F(v_R/3) + F(2v_R/3) + F(v_R/6) + \\ F(v_R/2) + F(v_R/10) + F(v_R/5) + F(3v_R/10) + F(2v_R/5) \quad (7)$$

for a range of values of the sidereal rotation frequency $v_R$. The result is shown in Figure 2. The principal peak is found at $v_R = 14.30$, which is within the range $10 - 15 \text{ year}^{-1}$ that is our conventional search band for internal rotation frequencies [9,10]. This corresponds to a period of 25.54 days, which agrees with the "fundamental period" proposed by Bai [2]. (The next biggest peak is at 7.14, which is one half the frequency of the principal peak, so that it is related to the principal peak.)

Table 1. Comparison of frequencies (nu, in year-1) and periods (P, in days) as calculated from Equations (2) and (4), with $v_R = 14.30 \text{ year}^{-1}$ and as tabulated by Bai [2].

|  | l | m | Calc nu | **Bai nu** | Calc P | **Bai P** |
|---|---|---|---|---|---|---|
| S-type | 2 | 1 | 4.81 | **4.81** | 76.6 | **76** |
| S-type | 2 | 2 | 9.62 | | | |
| S-type | 3 | 1 | 2.41 | **2.39** | 153.3 | **153** |
| S-type | 3 | 2 | 4.81 | **4.81** | 76.6 | **76** |
| S-type | 3 | 3 | 7.22 | **7.16** | 51.1 | **51** |
| S-type | 4 | 1 | 1.44 | **1.42** | 255.4 | **257** |
| S-type | 4 | 2 | 2.89 | **2.85** | 127.7 | **128** |
| S-type | 4 | 3 | 4.33 | **4.35** | 85.1 | **84** |
| S-type | 4 | 4 | 5.77 | **5.80** | 63.9 | **63** |
| | | | | | | |
| E-type | 2 | 1 | 8.62 | | | |
| E-type | 2 | 2 | 17.24 | | | |
| E-type | 3 | 1 | 11.03 | **10.90** | 33.5 | **33.5** |
| E-type | 3 | 2 | 22.05 | | | |
| E-type | 3 | 3 | 33.08 | | | |
| E-type | 4 | 1 | 11.99 | | | |
| E-type | 4 | 2 | 23.97 | | | |
| E-type | 4 | 3 | 35.96 | | | |
| E-type | 4 | 4 | 47.95 | | | |

We find, from helioseismology data [11], that $14.30 \text{ year}^{-1}$ is the rotation frequency at normalized radius 0.72. This falls within the familiar tachocline that extends over the normalized-radius range 0.68 to 0.74.

We show in Table 1 the S-type and E-type r-mode frequencies (given by Equations (4) and (2), respectively) for $l = 2,3,4$, and the relevant values of $m$, and for the sidereal rotation frequency ($14.30 \text{ year}^{-1}$) inferred from Figure 2. (Note that these equations lead to the same frequencies for $l = 2, m = 1$ and for $l = 3, m = 2$, so that there are only seven independent estimates for the S-type frequencies.) One of the periodicities listed by Bai (with period 33.5 days) may be identified with an E-type oscillation for the same sidereal rotation frequency, and for $l = 3, m = 1$. This is, of course, the E-type periodicity corresponding to the principal Rieger periodicity. We see that these estimates agree very well with all eight of the periods listed by Bai [2].

We show in Figure 3 the oscillation frequencies shown in Figure 1, with the frequencies normalized to the inferred sidereal rotation frequency $14.30 \text{ year}^{-1}$. This display also shows that the eight periodicities listed by Bai [2] fit very well with those of r-mode oscillations (seven S-type and one E-type) as they might occur in the solar tachocline.

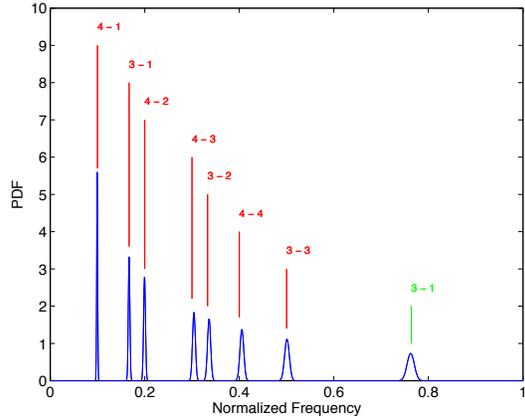

Figure 3. Representation of the Rieger and Rieger-related oscillations listed by Bai [2], together with the r-mode frequencies, with frequency normalized to the estimated sidereal rotation frequency, $v_R = 14.30 \text{ year}^{-1}$. S-type oscillations in red, E-type oscillation in green.

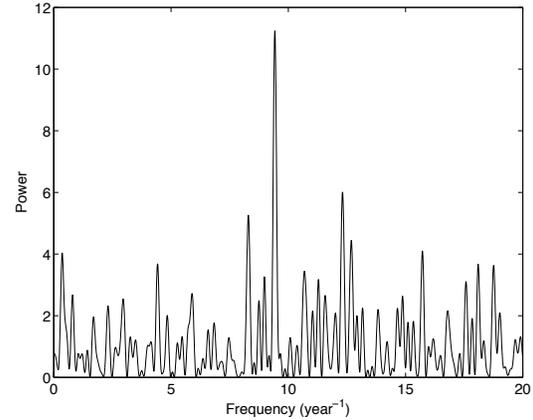

Figure 4. Power spectrum of Super-Kamiokande 5-day data, showing the prominent peak at $9.43 \text{ year}^{-1}$ with power $S = 11.24$.

## 4 . Super-Kamiokande Periodicities

The most precise measurements of the solar neutrino flux are those obtained by the Super-Kamiokande experiment [12]. A Lomb-Scargle analysis (that assigns a mean time to each time bin) of Super-Kamiokande data, as carried out by the Super-Kamiokande Consortium, does not yield persuasive evidence for any periodicity [13]. On the other hand, a likelihood analysis that takes account of the start-time and end-time of each bin and allows for a floating offset yields a significant periodicity with frequency 9.43 year[-1] and power $S = 11.24$, as shown in Figure 4 [14]. The results of a shuffle test show that the probability of finding this periodicity by chance, expressed as a function of the width of the search band, is 3.86 10[-4] per year[-1]. For instance, the probability of finding this periodicity by chance in the band 5 – 15 year[-1] is 0.4%. The 9.43 year[-1] frequency is a significantly lower frequency than would be expected for rotational modulation that has its origin in either the convection zone or the radiative zone.

We have also found that the power spectrum formed from Super-Kamiokande data contains five peaks that may be interpreted as E-type r-mode oscillations for $l = 2,3,4,5,6,$ and $m = 1$ and a sidereal rotation frequency of $13.97 \text{ year}^{-1}$ [15]. We show the relevant section of the power spectrum, referred to the appropriately normalized frequency, in Figure 5. This estimate corresponds to the rotation frequency at normalized radius 0.7, placing the origin of these oscillations also in the tachocline. The probability that these periodicities might have occurred by chance has been estimated to be 0.3% [15]. We see that the 3-1 S-type oscillation shows up in both Figure 3 (for the Bai-listed frequencies) and Figure 5 (for the Super-Kamiokande data).

We find that the S-type frequency for $l = 2, m = 2$ has the value 9.43 year[-1] for a sidereal rotation frequency of 14.15 year[-1], which also falls well within the range appropriate for the tachocline. Hence we see that the principal periodicity in Super-Kamiokande data (with frequency 9.43 year[-1]) may be attributable to an r-mode oscillation in the tachocline.

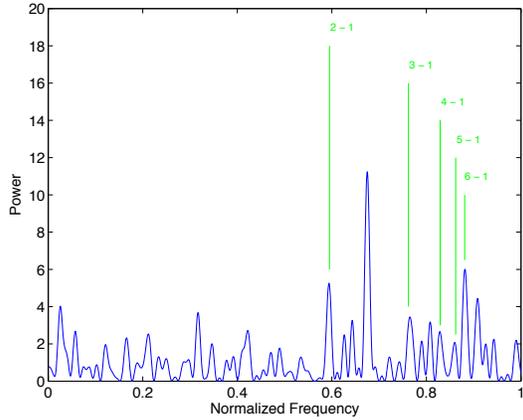
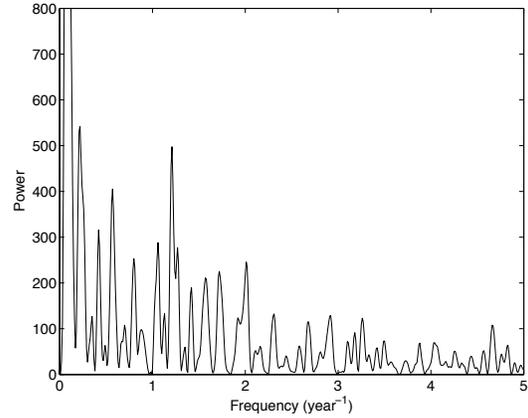

Figure 5. Section of the power spectrum of Super-Kamiokande 5-day data, showing the five E-type r-mode oscillations corresponding to $m = 1, l = 2,3,4,5,6$, with frequency normalized to the estimated sidereal rotation frequency, $v_R = 13.97$ year$^{-1}$.

Figure 6. Section of the power spectrum of Mt Wilson diameter data.

## 5. Periodicities in Mt Wilson Solar Diameter Measurements

We review briefly an analysis of 39,024 measurements of the solar diameter made at the Mt Wilson Solar Observatory over the years 1968 to 1998 [7]. The power spectrum formed from these measurements is shown in Figure 6. We have found that this power spectrum contains eight peaks that may be interpreted as S-type r-mode oscillations for $l = 2,3,4,5,6,7,8,10$ and $m = 1$ and a sidereal rotation frequency of 12.08 year$^{-1}$.

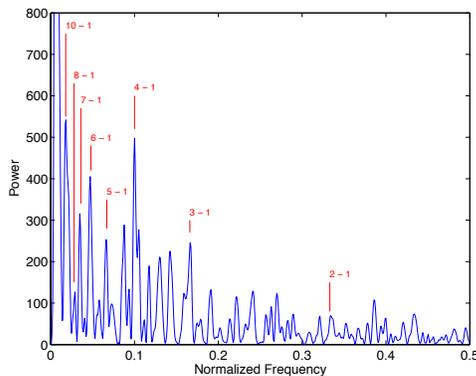

Figure 7. Section of the power spectrum formed from Mt Wilson Solar Diameter measurements, showing the eight S-type r-mode oscillations corresponding to $m = 1, l = 2,3,4,5,6,7,8,10$ and 12.08 year$^{-1}$.

These results are shown in Figure 7, in which the frequency has been normalized with respect to the inferred sidereal rotation frequency of 12.08 year$^{-1}$. We see that there is excellent correspondence between the r-mode frequencies and peaks in the power spectrum. The probability that these periodicities might have occurred by chance has been estimated, by means of the shuffle test, to be about $10^{-6}$ [7].

This estimate of the rotation rate is significantly lower than the range of sidereal rotation frequencies, as determined by helioseismology [6], for equatorial sections of either the convection zone (14.6 – 14.8 year$^{-1}$) or the radiative zone (13.5 – 13.9 year$^{-1}$).

## 6. Periodicities in Nuclear Decay Measurements made at the Lomonosov Moscow State University (LMSU)

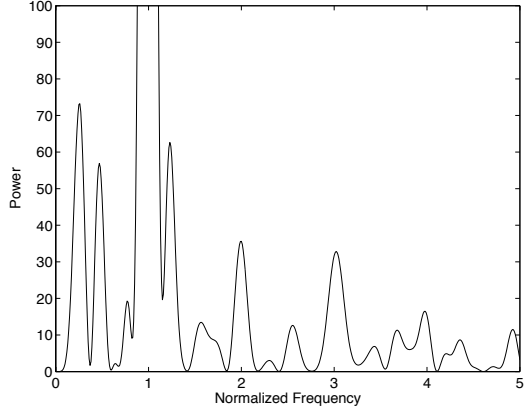

Figure 8. Section of the power spectrum of LMSU decay data.

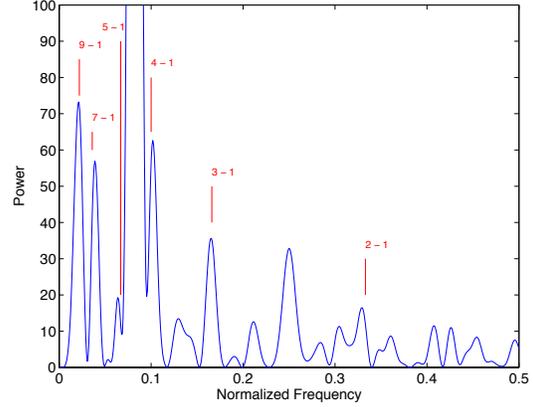

Figure 9. Section of the power spectrum formed from LMSU nuclear decay measurements, showing the six S-type r-mode oscillations corresponding to $m = 1, l = 2,3,4,5,7,9$, 12.08 year$^{-1}$.

Table 2. Comparison of S-type frequencies (nu, in year$^{-1}$), as calculated from Equation (4) with $\nu_R$ = 12.08 year$^{-1}$, with peaks in Mt Wilson and LMSU power spectra.

| | | Calc | Mt W | Mt W | LMSU | LMSU |
|---|---|---|---|---|---|---|
| l | m | nu | nu | S | nu | S |
| 2 | 1 | **4.03** | **4.04** | *69.49* | **3.98** | *16.49* |
| 3 | 1 | **2.01** | **2.01** | *245.62* | **2.00** | *35.60* |
| 4 | 1 | **1.21** | **1.21** | *497.52* | **1.23** | *62.62* |
| 5 | 1 | **0.81** | **0.80** | *253.02* | **0.77** | *19.28* |
| 6 | 1 | **0.58** | **0.57** | *405.12* | | |
| 7 | 1 | **0.43** | **0.42** | *315.54* | **0.47** | *56.90* |
| 8 | 1 | **0.34** | **0.35** | *126.76* | | |
| 9 | 1 | **0.27** | | | **0.26** | *73.23* |
| 10 | 1 | **0.22** | **0.22** | *541.96* | | |

We now review a recent similar analysis of 2,350 measurements of the $^{90}$Sr-$^{90}$Y decay process acquired over the years 2002 to 2009 at the Lomonosov Moscow State University [16]. The power spectrum formed from these measurements, shown in Figure 8, was found to contain six peaks that may be interpreted as S-type r-mode oscillations for $l = 2,3,4,5,7,9$ and $m = 1$ and a sidereal rotation frequency of 12.08 year$^{-1}$. (This is precisely the same frequency as found in our analysis of Mt Wilson diameter data.) We show in Figure 9 the power as a function of frequency normalized with respect to the inferred sidereal rotation frequency of 12.08 year$^{-1}$. The probability that these periodicities might have occurred by chance was estimated to be completely negligible. [16] The frequencies and powers of the relevant peaks in the Mt Wilson and LMSU power spectra are listed in Table 2, together with the corresponding r-mode frequencies for a sidereal rotation rate of 12.08 year$^{-1}$.

## 7. Resonant Spin Flavor Precession (RSFP)

We now consider the possibility that the modulation of the solar neutrino flux, as measured by the Super-Kamiokande experiment, is due to the RSFP (Resonant Spin Flavor Transition) process. [17-20] This is a mechanism by which neutrinos of one flavor, in traveling through matter permeated by a transverse magnetic field, would be converted to a different flavor. Since nuclear processes in the solar core produce electron neutrinos, and since Super-Kamiokande detects only electron neutrinos, the effect of the RSFP process would be to effect a reduction in the measured flux. This reduction would be in addition to that due to the MSW (Mikheyev, Smirnov, Wolfenstein) process, which leads to flavor changes by a process that does not involve magnetic field. [21,22] More realistic calculations would, of course, take account of both the MSW and RSFP processes.

There are two conditions to be satisfied in order for the RSFP process to occur. [19] The first is the *Resonance Condition*, which requires that

$$G_F \sqrt{2}(N_e - N_n) = \frac{\Delta(m^2)}{2E}, \quad (8)$$

where $\Delta(m^2)$ is the difference of $m^2$ between the two flavors of neutrinos, E is the neutrino energy, and $N_e$ and $N_n$ are the number densities of electrons and neutrons, respectively. If we measure mass and energy in eV and length in cm, the Fermi constant has the value

$$G_F = 10^{-37.03} \, eV \, cm^3. \quad (9)$$

The second condition is the *Adiabaticity Condition*, which requires that

$$H > \frac{G_F \sqrt{2}(N_e - N_n)}{4(\mu/\mu_B)^2 \mu_B^2 B^2} \quad (10)$$

where H is the scale height of the atmosphere, $\mu$ is the neutrino magnetic moment, $\mu_B$ is the Bohr magnetic moment, and the magnetic field strength *B* is measured in gauss. In these units, the Bohr magneton has the following value:

$$\mu_B = 10^{-7.23} \, eV \, G^{-1}. \quad (11)$$

Since the scale height is known for a given region of the solar interior, we choose to express Equation (10) as follows:

$$(\mu/\mu_B)B > 10^{-11.51}(N_e - N_n)^{1/2} H^{-1/2}. \quad (12)$$

The outer tachocline is centered on 0.7*R*, where *R* is the solar radius. At this radius, $N_e - N_n \approx 10^{23.0} \, cm^{-3}$. [23] Adopting 5 MeV ($10^{6.7}$ eV) as the typical energy of neutrinos detected by Super-Kamiokande, we find from the Resonance Condition [Equation (8)] that $\Delta(m^2) = 10^{-6.9} \, eV^{-2}$. This agrees with the estimate obtained by Das et al. [20] for a similar model.

We now consider the hypothetical inner tachocline and assume, as an example, that it is located at 0.25 R. At that radius, $N_e - N_n \approx 10^{24.9} \, cm^{-3}$. [23] Combining this figure with our estimate of $\Delta(m^2)$, we infer that neutrinos that satisfy the resonance condition (so that they can be modulated by the magnetic field) have an energy of about $10^{4.9}$ eV (80 keV).

In the outer tachocline, where H = $10^{8.8}$ cm, the Adiabaticity Condition [Equation (12)] becomes

$$(\mu/\mu_B)B > 10^{-4.4} \text{ for the outer tachocline.} \quad (13)$$

If, following the study by Weber, Fan, and Miesch [24], we adopt 80 kG ($10^{4.9}$ G) as the magnetic field strength in the outer tachocline, the Adiabaticity Condition yields the inequality $\mu/\mu_B > 10^{-9.3}$, which is a somewhat higher value than that ($\mu/\mu_B = 10^{-10}$) usually adopted. [20] If we adopt $\mu/\mu_B = 10^{-10}$, Equation (13) requires that $B > 10^{5.6} \, G$, which is significantly higher than current estimates of the magnetic field strength in that region of the solar interior.

We now consider the implications of the Adiabaticity Condition in the proposed inner tachocline. Assuming a location at 0.25 R, where H = $10^{9.7}$ cm, we find that

$$(\mu/\mu_B) B > 10^{-3.9} \text{ for the inner tachocline.} \quad (14)$$

For $\mu/\mu_B = 10^{-10}$, this condition would be met with a field strength of $10^{6.1}$ G, i.e. 1.3 $10^6$ G

## 8. Discussion

This article has been stimulated by recent evidence that some nuclear decay rates are not constant, and that the Sun is the cause—or one of the causes—of variability. The patterns of variability include an annual variation that is due in part (but not completely) to the eccentricity of the Earth's orbit [25-27]; variations with frequencies of order 10 - 12 year$^{-1}$ that appear to be due to internal solar rotation [28-30]; and variations with periods of the order of months that we suggest may be attributed to r-mode oscillations [31]. Examination of this topic raises problem of both solar physics and nuclear physics.

We saw in Section 3 that all eight of the Rieger-type periodicities in solar-activity data, as listed by Bai [2], may be understood as r-mode oscillations. Seven of the eight are S-type oscillations with frequencies given by Equation (4), and one is an E-type oscillation with frequency given by Equation (2). We infer from Equation (3) that the S-type periodicities may be attributed to the interaction of an r-mode with a non-uniform magnetic flux system that has components corresponding to several values of $m_S$. These may all be due to a localized magnetic region, since a delta function generates oscillations of all wave numbers. Possibly the interaction of an r-mode oscillation with magnetic flux in the tachocline is such as to lead to the eruption of a new active region or to the perturbation of an existing active region. The E-type periodicity, on the other hand, corresponds to an oscillation for which $m_S = 0$: that is, the r-mode is having a direct influence on solar activity.

There is no obvious reason why the only r-modes to be excited are those listed in Table 1, except perhaps that low-frequency components grow more readily than high-frequency components. With these formulas as a guide, we may find that solar activity is influenced by other r-mode oscillations, but not as obviously.

The analysis of Super-Kamiokande data in Section 4 indicates that r-mode oscillations can influence the solar neutrino flux. There are two well-known mechanisms that can influence neutrinos: the MSW (Mikheyev, Smirnov, Wolfenstein) mechanism [21,22], and the RSFP (Resonant Spin-Flavor Precession) mechanism [17-20]. According to MSW theory, spatial variations in density could lead to variations in the neutrino flux. However, one does not expect there to be any significant density inhomogeneity (apart from the radial variation) in the tachocline, which we believe to be the location of the modulation of neutrinos detected by Super-Kamiokande. According to RSFP theory, the neutrino flux can be modulated by a transverse magnetic field, coupled with a nonzero mass density. As we see in Section 7, it appears that the relevant conditions can be met in the solar interior.

We see in Sections 5 and 6 that there appears to be evidence that r-mode oscillations also occur in a region with a much lower rotation rate than that of either the convection zone or the radiative zone. We propose that this region is located below the radiative zone, for which helioseismology does not yet yield relevant information. As we see in Section 7, it is possible to attribute the relevant variations in decay rates to the RSFP effect if the solar core rotates sufficiently slowly and if the magnetic field strength is of order 1 MG. The diameter variations discussed in Section 5 are presumably due to some form of wave (possibly magneto-hydrodynamic) that propagates from the neighborhood of the core to the photosphere. The large decrease in density during propagation is likely to bring about a large increase in the displacement amplitude. However, we have made the simplifying assumption of considering the RSFP independently of the MSW process. We should also bear in mind that there are three different flavors of neutrinos (and a possible sterile neutrino). Different experiments may involve different combinations of these neutrinos.

It is notable that the oscillations evident in activity data, Super-Kamiokande data, diameter data, and LMSU decay data, appear to have their origin in two localized regions, one of which may be identified with the known tachocline, and the other of which we suggest may be identified with a hypothetical inner tachocline that separates the core from the radiative zone. This raises the question as to why r-mode oscillations should be preferentially excited in a tachocline. We offer (in the Appendix) the suggestion that r-mode oscillations are unstable in the presence of a radial gradient in the rotation rate, so that they may be expected to have significant amplitudes in such regions.

The sidereal rotation rate inferred from the Rieger and related oscillations (about 14 year$^{-1}$) may be identified with the rotation rate in or near the center of the tachocline. We have found that the sidereal rotation rate inferred from the Mt Wilson diameter data and the LMSU decay data is close to 12 year$^{-1}$. If this is the center of a hypothetical inner tachocline, this rotation rate may be the mean of the rotation rate of the radiative zone and the rotation rate of the core. The sidereal rotation rate of the radiative zone is believed to be about 13.5 year$^{-1}$. If these assumptions are correct, we may infer that the sidereal rotation rate of the core may be about 10.5 year$^{-1}$, which would correspond to a synodic rotation rate of 9.5 year$^{-1}$. Hence it is possible the salient periodicity in Super-Kamiokande solar neutrino data (at 9.43 year$^{-1}$) has its origin in the rotation of the solar core.

The physics problem posed by anomalous nuclear decays may be examined from the viewpoint of communication theory. This viewpoint leads one to ask three questions: (a) What is the channel of communication? (b) How is a signal injected into that channel? (What is the transmitter?) and (c) How is the signal extracted from that channel? (What is the receiver?)

We know something about (c): Some radioactive nuclei are receiving some kind of signal that influences their decay rates.

We also know something about (a): Some process or activity in the Sun is responsible for whatever emanation influences radioactive nuclei.

This leaves (b) as the crucial question. What is the nature of the radiation that is emitted by the Sun and which influences radioactivity? This radiation must satisfy the following two requirements:

(i) There must be a mechanism, which occurs in the solar interior, for modulating that radiation.

(ii) That modulation must survive transmission from the solar interior to (and perhaps through) the Earth.

These two requirements point to neutrinos as the prime suspect for the radiation mechanism. It is known that neutrinos have an exceedingly small cross-section for normal interaction with matter, and it is known that neutrinos created in the solar interior can reach detectors on Earth. The "modulation" of the neutrino flux could be simply an "amplitude" modulation of the flux. An alternative is modulation of the flavor of the neutrinos.

Fluctuations in the nuclear reaction rate could lead to amplitude modulation. Inhomogeneity in the burning, coupled with the MSW effect, could give rise to such modulation. However, since r-mode oscillations are almost isobaric, it is difficult to see how they could influence the nuclear burning rate.

On the other hand, flavor modulation can be brought about by magnetic field via the RSFP process, which we examined in Section 7. Nuclear reactions produce electron-flavor neutrinos. Although there is at this time no known process by which neutrinos of any flavor can influence nuclear decays, it is reasonable to suppose that, if such a process can be found, its efficacy is likely to depend upon the neutrino flavor.

The results reported in this article raise a number of interesting questions that can best be pursued by similar analyses of different data sets. We could, for instance, search for evidence of r-mode oscillations in solar-diameter measurements acquired by the MDI and RHESSI experiments, and possibly in measurements of the total solar irradiance (TSI). If such measurements confirm such oscillations in photospheric data, this will raise the theoretical problem of identifying the mechanism by which oscillations in or near the solar core can influence processes in the photosphere.

Oscillations in Super-Kamiokande data and in nuclear decay data are suggestive of the influence of solar neutrinos. It would therefore be interesting to carry out similar analyses of other neutrino data sets, including data acquired by the Super-Kamiokande experiment after 2001. It will also be interesting to examine nuclear-decay data, other than the LMSU data analyzed in Section 6, to search for similar evidence indicative of r-mode oscillations.

If further data analyses support the suggestion that neutrinos can influence nuclear decay rates, one will face the challenging problem of understanding the relevant mechanism.

The work of EF was supported in part by U.S. DOE contract No. DE-AC02-76ER071428. We are indebted to Taeil Bai and Joao Pulido for helpful discussions concerning this project. The views expressed in this paper are those of the authors and do not reflect the official policy or position of the U.S. Air Force, U.S. Department of Defense, or the U.S. Government.

**Appendix A. Discussion of a Possible Kelvin-Helmholtz Instability of r-Modes in a Tachocline**

It is well known that a latitudinal gradient in angular velocity can lead to instability in a rotating star [4, 32]. We here explore conceptually the possibility that a similar instability may be caused by a radial gradient of angular velocity.

It is well known that a gradient in flow velocity can lead to instability of a fluid (the Kelvin-Helmholtz Instability) [33] and that a superposition of streams with different velocities can lead to the two-stream instability [34]. In general, if two otherwise identical physical systems have excitations $\chi_1$ and $\chi_2$, and if a coupling of the systems would lead to an excitation of $\frac{1}{2}(\chi_1 + \chi_2)$, and if the energy is proportional to $\chi^2$, then the coupling will lead to a release of energy proportional to $(\chi_1 - \chi_2)^2$. An r-mode oscillation is such an excitation, so one might expect that the coupling of two neighboring r-mode oscillations will be unstable.

In a uniformly rotating fluid, the frequency of r-mode oscillations is given, to good approximation, by
$$\omega = m\Omega - \frac{2m\Omega}{l(l+1)} , \tag{A.1}$$
where $\Omega$ is the angular rotation rate. Since this expression does not involve the radial index *n*, one could (to this approximation) consider an r-mode oscillation that is confined to a thin spherical shell. If we now consider two adjacent shells with different angular velocities, we should expect that any coupling of the two excitations would lead to instability.

As an example of such a system, we now consider a weak coupling of two adjacent spherical shells with different angular velocities.

It is convenient to rewrite Equation (A.1) as
$$\omega = mh\Omega , \tag{A.2}$$
where
$$h = 1 - \frac{2}{l(l+1)} . \tag{A.3}$$

We consider perturbations of the fluid of the form
$$\chi = \tilde{\chi} e^{i(m\phi - \omega t)}, \quad (A.4)$$
where $\chi$ may be considered to be the oscillatory part of the velocity of a fluid element.

The wave equation corresponding to the heuristic model, for which the dispersion relation is Equation (A.2), is seen to be
$$\frac{\partial \chi}{\partial t} + h\Omega \frac{\partial \chi}{\partial \phi} = 0. \quad (A.5)$$
Hence an r-mode oscillation may be viewed conceptually as an excitation of a system that satisfies this wave equation.

We now consider r-mode excitations in two adjacent shells with angular velocities $\Omega_1$ and $\Omega_2$, and we suppose that there is a weak coupling between these shells, leading to the following pair of coupled wave equations:
$$\frac{\partial \chi_1}{\partial t} + h\Omega_1 \frac{\partial \chi_1}{\partial \phi} + k(\chi_2 - \chi_1) = 0,$$
$$\frac{\partial \chi_2}{\partial t} + h\Omega_2 \frac{\partial \chi_2}{\partial \phi} + k(\chi_1 - \chi_2) = 0. \quad (A.6)$$
The coupling may have its origin in an effective friction due to small-scale turbulence, or it could be due to the fact that the eigen-modes are not strictly independent of the radial index *n*, so that the quasi-eigen-modes formed from a combination of *n* values have a finite, but small, radial extent.

If we now consider waves of the form of Equation (A.4), we obtain the following equations for the complex amplitudes:
$$-i\omega\tilde{\chi}_1 + imh\Omega_1\tilde{\chi}_1 + k(\tilde{\chi}_2 - \tilde{\chi}_1) = 0,$$
$$-i\omega\tilde{\chi}_2 + imh\Omega_2\tilde{\chi}_2 + k(\tilde{\chi}_1 - \tilde{\chi}_2) = 0. \quad (A.7)$$
On forming the determinant of this pair of equations, we obtain the following dispersion relation for the coupled system:
$$\omega = mh\left(\frac{\Omega_1 + \Omega_2}{2}\right) + ik \pm \left[m^2 h^2 \left(\frac{\Omega_1 - \Omega_2}{2}\right)^2 - k^2\right]^{1/2}. \quad (A.8)$$
For any values of the parameters, the imaginary part of $\omega$ is found to be positive, informing us that the wave grows in amplitude. In the weak-coupling approximation, that $k$ is small compared with $mh|\Omega_1 - \Omega_2|$, the values are, approximately,
$$\omega = mh\Omega_1 + ik, \text{ and } mh\Omega_2 + ik, \quad (A.9)$$
showing that the growth rate is given by the coupling coefficient *k*.

This result suggests that if there is a gradient in angular velocity (of either sign), an r-mode in that location is unstable and will grow in amplitude. This suggests that r-modes will grow to finite amplitude in a tachocline.

**Appendix B: Comments on Evidence for Variability of Nuclear Decay Rates**

Some authors have expressed concerns with respect to our articles (notably our first article [26]) presenting evidence for the variability of certain nuclear decay rates. We discuss some of these concerns briefly in this appendix.

Semkow et al. [35] pointed out that measurements of radiative-decay products can be influenced by the typical annual temperature variation since an increase in temperature can lead to a decrease in density, which in turn leads to a smaller attenuation of decay products such as gamma rays. They suggested that the annual variation of $^{226}$Ra decay rates measured in the PTB (Physikalisch-Technische Bundenanstalt) experiments [36] could have this explanation. In response, Jenkins et al. [37] point out that the PTB experiment incorporated a sealed chamber so that the density of gas within the chamber would not be influenced by variations in the ambient temperature. Concerning the BNL (Brookhaven National Laboratory) experiments [38], the experimenters had themselves considered the possibility that the annual oscillations in their measurements might have been caused by annual environmental oscillations, but concluded that "in order to produce the variations of +/- 3 standard deviations, the large humidity changes would have to be combined with temperature variations over a range of at least +/- 5 deg F, which is larger than the probable actual range." [38]

Cooper [39] examined data from the radioisotope thermoelectric generators (RTGs) on board the Cassini spacecraft that was launched in 1997 and reached Saturn in 2004. During this period, the distance of the spacecraft from the Sun varied over the range 0.7 AU to 1.6 AU. Cooper found no evidence for a dependence of the output of the RTGs on the spacecraft-Sun distance, and claimed that this calls into question evidence for the annual variation in the $^{226}$Ra decay rate measured in the PTB experiment. In response, Jenkins et al. [36] point out that they had only claimed to find evidence for decays which involve the beta process, making no claim concerning decays involving only alpha processes. Krause et al. [40] have prepared a more detailed response, pointing out that RTGs derive their power from $^{238}$Pu, which decays by alpha emission to $^{234}$U, which in turn decays by alpha emission to $^{230}$Th. By contrast, the decay chain of $^{226}$Ra, which was discussed in Jenkins et al. [26], involves both alpha and beta decays.

Schrader [41] has recently published a sub-set of the PTB data (for the interval 1990 to 1996) and claims that the ratios of pairs of measurements do not exhibit annual variations. However, Schrader has generously made that subset available to us, and we find that some of the nuclides (notably $^{133}$Ba) do in fact exhibit an annual modulation.

Norman et al. [42] have reported that they had analyzed measurements of the decay rates of several nuclides, notably $^{22}$Na/$^{44}$Ti, and found no evidence of an annual oscillation. However, their analysis assumed a fixed amplitude and a fixed phase for a possible annual oscillation. O'Keefe et al. [43] have found that a more flexible analysis, such as a Lomb-Scargle power-spectrum analysis, in fact yields evidence in the Norman $^{22}$Na/$^{44}$Ti data for an annual oscillation, significant at the 1% level.

de Meijer, Blaauw, and Smit [44] have recently reported the results of an experiment in which they obtained high-precision measurements of gamma-ray count rates during reactor-on and reactor-off periods to investigate the possible influence of antineutrinos on the nuclear-decay rates of $^{152}$Eu, $^{137}$Cs, $^{54}$Mn, and $^{22}$Na. This experiment showed no evidence for such an influence. However, these results are not in conflict with the results of our analyses for two reasons: (a) The energies of antineutrinos from reactors are typically in the range 3-4 MeV whereas only a very small fraction of solar neutrinos are in this energy range. The most abundant solar neutrinos are produced by p-p reactions, with energies up to only 400 keV. (b) The radiation that causes variations in nuclear decay rates is unknown: it could be due to electron, muon or tau neutrinos, or antineutrinos, or some other form of radiation yet unknown.

Bellotti et al. [45] have recently reported the result of measurements of the activity of a $^{137}$Cs source, as determined by an experiment in the Gran Sasso Laboratory. They report that "no signal with amplitude larger than $9.6 \times 10^{-5}$ (at 95% C.L.) has been detected," concluding that this result is "in clear contradiction with previous experimental results and their interpretation as indication of a novel field (or particle) from the Sun." In reviewing the case for variability, Bellotti et al. refer to articles by Jenkins et al. [25], Fischbach et al. [46], Parkhomov [47,48], and Javorsek et al. [49]. However, none of these articles cites decay rates for $^{137}$Cs. We find that the PTB measurements of the

$^{137}$Cs decay rate show no evidence of an annual oscillation, in agreement with the Bellotti result. However, PTB measurements of $^{133}$Ba and of $^{226}$Ra decay rates do show evidence of annual oscillations. Jenkins et al. [50] have prepared a more detailed discussion of the Bellotti article.

To sum up: It is clear that different nuclides behave differently. We have found no evidence to date that any alpha decays are variable. It appears that some—but not all—beta decay rates are variable. The mechanism is presently unknown.